\documentstyle[aps]{revtex}
\def\be {\begin{equation}}
\def\ee {\end{equation}}
\def\ba {\begin{eqnarray}}
\def\ea {\end{eqnarray}}

\def\d  {\delta}

\def\m  {\mu}

\def\o  {\omega}

\def\p  {\pi}
\def\P  {\Pi}

\def\la {\label}
\def\le {\left}
\def\ri {\right}
\def\pa {\partial}
\def\f {\frac}

\def\no {\noindent}
\def\bi {\begin{itemize}}
\def\ei {\end{itemize}}

\def\vs {\vspace}

\def\da {\dagger}
\def\ha {\hat}
\def\car {{\cal R}}

\begin{document}
\draft
\title{Quantum Mechanical Spectra of Charged Black Holes}
\author{ Saurya Das}
\address{Department of Mathematics and Statistics,} 
%\address{}
\address{University of New Brunswick,}  
\address{Fredericton, New Brunswick - E3B 5A3, CANADA}

%\address{ 
\address{ EMail: saurya@math.unb.ca   }
\author{P. Ramadevi, U. A. Yajnik, A. Sule}
\address{Department of Physics,}
\address{Indian Institute of Technology Bombay,} 
\address{Mumbai - 400 076, INDIA}
\address{EMail: ramadevi,yajnik@phy.iitb.ac.in}
\maketitle

\begin{abstract}
It has been argued by several authors, using different
formalisms, that 
the quantum mechanical spectrum of black hole horizon
area is discrete
and uniformly spaced. Recently it was shown that two
such approaches,
namely the one involving quantisation on a reduced
phase space, 
and the algebraic approach of Bekenstein and
Gour are equivalent for 
spherically symmetric, neutral black holes
(hep-th/0202076). That is,  
the observables of one can be mapped to those of the
other.  
Here we extend that analysis to include charged black
holes. 
Once again, we find that the ground state of the black
hole is 
a Planck size remnant. 

\end{abstract}

%%%%%%%%%%%%%%%%%%%%%%%%%%%%%%%%%%%%%%%%%%%%%%%%%%%%%%%%%%%%%%%%%%%%%%%%%%%
\vs{.6cm}

Black holes, in addition to being fascinating objects
in our universe, 
serve as theoretical laboratories where many
predictions of 
quantum gravity can be tested. It is well known that
quantum mechanics 
plays a crucial role in many phenomena involving black
holes, e.g. 
Hawking radiation and Bekenstein-Hawking entropy. Thus
it is important
to explore the quantum mechanical spectra of 
black hole observables such as horizon area, charge
and angular 
momentum. It has been argued by various authors, using
widely 
different approaches, that the spectra of above
observables are discrete 
\cite{bek,bm,bekgour,kogan,bk,bdk,Berezin,kastrup,louko,kastrup_strobl,VW,garattini,MRLP,poly}.
In particular, the horizon area of a black hole has
been shown to have 
a uniformly spaced spectrum. Though the spectrum found
in \cite{others} is not strictly uniformly spaced, 
it is effectively equally spaced for large areas.
  
As distinct as they may seem, since the different
approaches attempt
to address similar questions and predict similar
spectra, 
it is expected that there is an underlying connection 
between them. 
In \cite{dry} we examined this issue for
two of the above 
approaches, namely that advocated in \cite{bk,bdk} and
that in 
\cite{bm}, for black holes which are {\it spherically
symmetric}
and {\it neutral}. A direct mapping of the operators in the
two approaches was found in that article, which was essential
to get a physical interpretation of the abstract operators 
in the second approach. Moreover, we showed that the exact 
`quantum' of horizon area (which turns out to be the square
of Planck length) cannot be determined without this mapping. 

In view of the above results, it is important to see how robust the
results in the two approaches and the mapping between the two are. 
In this article, we try to address this question by relaxing the 
assumption that the black hole is uncharged, and consider black
holes carrying an electric (or magnetic)  charge instead. 
The two approaches whose underlying connections 
we will study  are:
\begin{enumerate}
\item The reduced phase space quantisation of
spherically symmetric black hole configurations of gravity \cite {bk,bdk}.
To make it amenable to quantisation, a
canonical transformation is 
performed. We will call this {\it Approach I}.
\item An algebra of black hole observables postulated
by Bekenstein and Gour \cite {bm,bekgour} 
giving uniformly spaced area spectrum.
We will call this {\it Approach II}. 
\end{enumerate}

It was shown in \cite{dry} that for {\it neutral} black holes, 
the observables of one approach can be rigorously
mapped to those of
the other. Here, we show that the mapping can be extended 
to incorporate charged black holes. We also show
that inclusion of charge leaves
the important features of the spectrum unchanged,
namely that the spectrum is still discrete and uniformly spaced.
However, a new 
quantum number enters the picture, 
associated with the $U(1)$ charge, and the
corresponding horizon area
now depends linearly on two quantum numbers, instead
of just one. 

In Approach II, one of the starting points is the 
assumption that horizon area is an adiabatic
invariant, and from
Bohr-Sommerfeld quantisation rule which stipulates
that adiabatic
invariants must be quantized \cite{pauling}, it
follows that the
area spectrum is discrete and uniformly spaced. 
In Approach I, on the other hand, a result which is
similar to the above
conjecture, was
explicitly proven for spherically symmetric black
holes which are away
extremality. In particular, it was shown that the
horizon area above extremality
is an adiabatic invariant. We shall return to this
issue later. 

First we will briefly review the two methods. It
follows from the analysis 
of \cite{dk,mk} that the dynamics of
static spherically symmetric charged black hole 
configurations in {\it any} classical theory of gravity in 
$d$-spacetime dimensions is governed by an effective action of the
form
\be
I = \int dt \le( P_M {\dot M} + P_Q {\dot Q} - H(M,Q)
\ri) 
\ee
where $M$ and $Q$ are the  mass  and the charge of the 
black hole
respectively and $P_M, P_Q$ the corresponding
conjugate momentum. This is essentially a consequence
of the no-hair theorem.
The boundary conditions imposed on these variables
are those of \cite{lw,shelemy}.
It can be shown that $P_M$ has the physical
interpretation of 
asymptotic Schwarzschild time difference between the
left and right 
wedges of a Kruskal diagram
\cite{kuchar,thiemann,gkl}. 
Note that $H$ is independent
of $P_M$ and $P_Q$, such that from Hamilton's
equations, $M$ and $Q$ 
are constants of motion. 

Now to explicitly incorporate the thermodynamic 
properties of these black holes,  
%restrict ourselves to black holes (and
%simultaneously exclude all other
%spherically symmetric configurations), 
motivated by Euclidean 
quantum gravity \cite{euclidean}, we assume that the
conjugate
momentum $P_M$ is periodic with period equal to  
inverse Hawking temperature. That is,
\be
P_M \sim P_M + \f{1}{T_H(M,Q)} ~~.
\la{period}
\ee
Similar assumptions were made in the past using 
different arguments \cite{kastrup,louko,kastrup_strobl}.
Note that the above identification implies that the
the 
$(M,P_M)$ phase subspace has a wedge removed from it,
which makes
it difficult, if not impossible to quantise on the
full phase-space.
Thus, one can make a canonical transformation 
$(M,Q,P_M,P_Q) \rightarrow (X,Q,\P_X,\P_Q)$, which
on the one hand `opens up' the phase space, and on the
other hand, 
naturally incorporates the periodicity (\ref{period})
\cite{bk,bdk}:
\ba
X &=& \sqrt{\f{A-A_0}{4\p G_d}} \cos \le( 2\p P_M T_H \ri) 
\la{ct1} \\
\P_X &=& \sqrt{\f{A-A_0}{4\p G_d}} \sin \le( 2\p P_M T_H \ri) 
\la{ct2}\\
Q &=& Q \la{ct3} \\
\P_Q &=& P_Q + \Phi P_M + \f{\left(d[A_0(Q)]/dQ\right)
P_M T_H}{4G_d}    \la{ct4} 
\ea
where $A$ is the black hole horizon area, $G_d$ the
$d$-dimensional
Newton's constant.  
Note that both $A$ and $T_H$ are functions of $M$ and
$Q$.  
$A_0(Q)$ is the value of area at extremality when the
mass of the black hole approaches its charge. For 
a $d$-dimensional Reissner-Nordstr\"om black hole, the
value
of $A_0(Q)$ is
\be
A_0(Q) = k_d Q^{(d-2)/(d-3)}
\la{a0}
\ee
where
$k_d=(1/4)(A_{d-2}/G_d)^{(d-4)/2(d-3)}(8\pi/(d-2)(d-3))^{(d-2)/2(d-3)}$
with $A_{d-2} = 2\pi^{(d-1)/2}/\Gamma( (d-1)/2)$ (area
of unit $S^{d-2}$). 
Also, $\Phi$ is the electrostatic potential at the
horizon 
and it will be treated as a $c$-number in the
following.
The validity of the first law of
black hole thermodynamics ensures that the above set
of transformations
is indeed canonical \cite{bk,bdk}. 
Squaring and adding (\ref{ct1}) and (\ref{ct2}), we
get:
\be
A_1 \equiv A - A_0(Q) = 4 \p G_d \le( X^2 + \P_X^2
\ri)~~.
\la{ham1}
\ee
The r.h.s. is nothing but the Hamiltonian of a simple
harmonic oscillator defined on the $(X,\P_X)$ phase
space with mass $\m$
and angular frequency $\o$ given by $\m=1/\o=1/8\p
G_d$. 
Upon quantisation, the `position' and `momentum'
variables are 
replaced by the operators:
\be
X \rightarrow {\ha X}~~~,~~~
\P_X \rightarrow {\ha \P_X} = - i  \f{\pa}{\pa X}~~,
\la{op1}
\ee
and the spectrum of the operator $A_1$ follows
immediately :
\be
{\rm Spec} \{\hat A_1\} \equiv a_n =  n \bar a +
a_{Pl}~~~~~,~~n=0,1,2, \cdots~~,
\la{spec1}
\ee
where $\bar a = 8\pi G_d = 8\p \ell_{Pl}^{d-2}$ is the
basic
quantum of area, $a_{Pl} = \bar a/2$ is its
`zero-point' value
( $\ell_{pl}$ is the $d$-dimensional Planck length).

To complete the analysis of the spectrum, we use the
following result from \cite{dk}:
$$ \delta P_Q = -\Phi~\d P_M + \delta \lambda~~,$$
where $\Phi$ is the electrostatic potential on the
boundary under
consideration, and variation refers to small change in
boundary conditions,
$\lambda$ being the gauge parameter at the boundary.
This in turn implies that for compact $U(1)$ gauge
group,
$ \chi \equiv e\lambda = e (P_Q + \Phi P_M)$ is
periodic
with period $2\pi$ (where $e$ is the fundamental unit of
electric charge).
Also, we saw earlier from thermodynamic arguments
that $\alpha \equiv 2\pi P_M T_H(M,Q)$
has period $2\pi$. In terms of these
`angular' coordinates, the momentum $\Pi_Q$ in
(\ref{ct4}) can be written as:
$$ \Pi_Q = \frac{\chi}{e}  + \frac{
\le(d[A_0(Q)]/dQ \ri)
\alpha}{8\pi G_d}~~.$$
Thus, the following identification must hold in the
$(Q,\Pi_Q)$ subspace:
\begin{eqnarray}
\left(Q, \Pi_Q \right) \sim
\left( Q, \Pi_Q + \f{2\pi n_1}{e}  + 
\f{n_2 \le(d[A_0(Q)]/dQ \ri) }{4 G_d}
\right)~~,
\label{qpq}
\end{eqnarray}
for any two integers $n_1,n_2$. Now, wavefunctions of
charge eigenstates
are of the form:
$$ \psi_Q(\Pi_Q) = \exp\left(i Q\Pi_Q \right)
~~,$$
which is single valued under the identification
(\ref{qpq}),
provided there exists another integer $n_3$ such that:
$$ n_1 \frac{Q}{e} + 
\f{n_2 Q \le(d[A_0(Q)]/dQ \ri) }{8\pi G_d}
= n_3~~.$$
Now, it can be easily shown  that
the above conditions is satisfied if and only if the
following
two quantisation conditions hold:
\ba
\f{Q}{e} &=& m ~~,~m = 0, \pm 1, \pm 2, \cdots 
\la{ch1} \\
\f{Q}{8\pi G_d} \left( d[A_0(Q)]/dQ \right) &=& p 
~~~,~ p=0,1,2,\cdots     \la{ch2}  
\ea
For $d$-dimensional 
Reissner-Nordstr\"om black holes, using the expression
for $A_0(Q)$ from
Eq.(\ref{a0}),  and combining (\ref{spec1}),
(\ref{ch1}) and (\ref{ch2}) 
we get its final area spectrum: 
\be
{\rm Spec}\{ \hat A \} = {\rm Spec}\{\hat A_0 + \hat
A_1 \} 
\equiv a_{nm} = \le[ 
n+ \le( \f{d-3}{d-2} \ri) p \ri] \bar a +
a_{Pl}~~~~~~~n,p=0,1,2, \cdots~~, \la{rnspec}
\ee
where $m$ and $p$ are related by eqns. (\ref {ch1},
\ref {ch2}).
Hawking radiation takes place when the black hole
jumps
from a higher to a lower area level, the difference in
quanta being
radiated away. The above spectrum shows that the black
hole does not 
evaporate completely, but a Planck size remnant is
left over at the 
end of the evaporation process. 
It may be noted that the periodic classical orbits in
the phase space under 
consideration admit of an adiabatic invariant. From
(\ref{spec1}), it
can be seen that the adiabatic invariant in this case
is:  
\be
\mbox{Adiabatic Invariant} = \oint \P_X dX  =
\f{A_1}{4G} ~~.
\la{adinvt}
\ee
Thus for $A \gg A_0(Q)$ (i.e. far from extremality),
the horizon area is indeed an adiabatic invariant 
(as conjectured in \cite{bm}). However, close to 
extremality, the above relation suggests that it
is the area above extremality which is an adiabatic
invariant. The advantage of relation (\ref{adinvt}) is that
on the one hand it is consistent with the discrete spectra
(\ref{rnspec}), and on the other hand, it ensures that
the extremality bound $A \geq A_0(Q)$ is automatically 
obeyed. 

The above result indicates that 
the relevant operator in the algebra of approach II for a 
generic non-extremal black hole is ${\hat A}_1$, which
along with the charge operator ${\hat Q}$, and the black hole
creation operator  
${\hat {\cal R}}_{nms_{nm}}$ forms a closed algebra 
(we follow the notation of \cite{bm}).
The operator ${\hat {\cal R}}_{nms_{nm}}$ creates a single
black hole state
from the vacuum $\vert 0 \rangle$ with $\hat A_1$
eigenvalue $a_n$ and
${\hat Q}$ eigenvalue $me$  in an internal quantum
state
$s_{nm}$:
\ba
{\hat {\cal R}}_{nms_{nm}}\vert 0 \rangle &=& \vert
nms_{nm} \rangle \\
\hat A_1 \vert nms_{nm} \rangle &=& a_n \vert nms_{nm}
\rangle\\
\hat Q \vert nms_{nm} \rangle &=& me \vert nms_{nm}
\rangle~.
\ea
We choose $s_{nm} \in \{0,1, \ldots k_{nm}-1 \}$
as in \cite {gour} such that the degeneracy of states
with 
same total area eigenvalue $a_{nm}$, obeys $\ln k_{mn}
\propto a_{nm}$. 
All these states have the same area and charge, which
ensures that
the Bekenstein-Hawking area law for black hole entropy
is obeyed.

We shall denote area above extremality in Bekenstein's
algebra
as $\hat A^{\prime}_1$ with eigenvalues 
$a^{\prime}_n$ such that the lowest eigenvalue is
$a^{\prime}_0 =0$. We will shortly see the relation
between the
operators $\hat A_1$ and $\hat A^{\prime}_1$ and their
respective
eigenvalues $a_n$ and $a^{\prime}_n$.
From symmetry, linearity, closure and gauge invariance
of area operator, 
the algebra satisfied by the charged black hole
operators
will be \cite{bm} :
\ba
{[} {\hat Q} , {\ha A^{\prime}_1}{]} &=& 0
~,\la{cl2}\\
{[}{\hat A^{\prime}_1}, {\hat {\cal R}}_{nm s_{nm}}]
&=& 
a^{\prime}_n{\ha {\cal R}}_{n ms_{nm}}~, 
\la{al1}\\
%{[} {\hat A^{\prime}_1} , {\ha {\cal
%R}}_{nms_{nm}}^\da ] &=& - a^{\prime}_n
%{\ha {\cal R}}_{nms_{nm}}^\da ~,\la{al2}\\
{[} {\hat Q} , {\ha {\cal R}}_{nms_{nm}} ] &=& me 
{\ha {\cal R}}_{nms_{nm}} ~,\la{cl1}\\
{[} {\ha \car}_{nms_{nm}} ,  {\ha \car}_{n'm's_{n'm'}}
] &=& 
\epsilon_{nn'm m'}^{n''m''} 
{\ha \car}_{n''m''s_{n''m''}}
~~~(\epsilon_{nn'mm'}^{n''m''} \neq 0~~
{\rm iff}~ a^{\prime}_n+a^{\prime}_{n'} = a^{\prime}_{n''}
~{\rm and}~ m+m'=m'')~,
\la{al3'}\\
{[} {\hat A^{\prime}_1}, {[} {\ha {\cal
R}}_{n'm's_{n'm'}}^\da, 
{\ha {\cal R}}_{nms_{nm}} ] ] &=& (a^{\prime}_n -
a^{\prime}_{n'})
{[} {\hat {\cal R}}_{n'm's_{n'm'}}^\da, {\hat {\cal
R}}_{nms_{nm}} ]~~
{\rm iff} ~a^{\prime}_n > a^{\prime}_{n'} ~.\la{al3} 
\ea
Eqn.(\ref {al3'}) implies that the black hole state
created by a 
commutator of two black hole creation operators,
$[{\ha \car}_{nms_{nm}}, 
{\ha \car}_{n'm's'_{nm}}]$, will be another
single black hole state (${\ha \car}_{n''m''s''_{nm}}
\vert 0 \rangle$) provided
$a^{\prime}_n + a^{\prime}_{n'} = a^{\prime}_{n''}$
and $m+m'= m''$.
Clearly, $\hat A^{\prime}_1$ is a positive definite
operator because 
the area above extremality cannot be negative.
Incorporating  
this, and adjoint relation of eqn. (\ref {al1}), we 
require the inequality condition $a^{\prime}_n >
a^{\prime}_{n'}$ 
in eqn. (\ref {al3}).
Clearly, the spectrum of the above algebra involves
both addition 
and subtraction of $\hat A^{\prime}_1$ eigenvalues
which is possible if and 
only if the $\hat A^{\prime}_1$ eigenvalues are 
equally spaced, i.e.,
$a^{\prime}_n = n \bar b$ where $\bar b$ is an unknown
positive
constant with dimensions of area. 

Now, it can also be seen that the above algebra
(\ref {cl2}- \ref {al3}) is unchanged under a constant
shift of the 
$\hat A^{\prime}_1$ operator. Allowing this
possibility, we re-define:
\be
\hat A^{\prime}_1 \rightarrow \hat A^{\prime}_1 + \bar
c \hat I 
\equiv \hat A_1
~,
\ee
where $\bar c$ is an arbitrary constant. The above
relation implies that the
eigenvalues $a^{\prime}_n$ and $a_n$ are related as
follows:
\be
a_n = a^{\prime}_n + \bar c~.
\ee
Equivalently, the lowest eigenvalue $a_0$ is non-zero,
$a_0= \bar c$.
Comparing the algebraic approach with reduced phase
approach, the constant 
$\bar c= a_{Pl}= 4 \pi \ell_{p\ell}^{d-2}$ and the
unknown area spacing
$\bar b = \bar a$  so that the $\hat A_1$ spectrum
is the same as eqn. (\ref {spec1}). This fixes the
spectrum $\{a_n\}$
of $\hat A_1$ uniquely. Therefore the spectrum of the
total area 
operator for the charged black hole takes the form
\begin{equation}
a_{nm} = a_n + f(m)~, \label {algar}
\end{equation}
where $f(m)$ corresponds to the contribution from
area at extremality $A_0(Q)$. 
%There does not seem to be
%any provision for studying the quantisation
%of the electromagnetic sector within the algebraic
%approach to derive eqns. (\ref {ch1},\ref {ch2}). 
%
In order to determine the exact form of $f(m)$, first
note that it has to be proportional to ${\bar a}$, 
on dimensional grounds. Secondly, the extremality 
bound for a charged black hole has to be satisfied,
at least for macroscopic black holes. This unambiguously 
establishes the factor of proportionality to be $(d-3)p/(d-2)$, 
such that:
%
%In order to make the area spectra (\ref {algar}),
%(\ref {spec1}) of the two approaches identical,
%we have to set for the $d$-dimensional
%Reissner-Nordstr\"om black hole
\begin{equation}
f(m) = \le( \f{d-3}{d-2}\ri) p~{\bar a}~,
\end{equation}
where $m$ and $p$ are implicitly related by eqns.
(\ref{ch1}) and
(\ref {ch2}). Thus, the area spectra (\ref{rnspec})
and (\ref{algar}) become identical.

Our next step is to find a realization of the
operators in
{\it Approach II} in terms of the fundamental degrees
of freedom
($M, \Pi_M, Q, \Pi_Q$) in {\it Approach I}. We proceed
in two
steps. First, we propose a representation of the
algebra
(\ref {cl2}-\ref{al3}) with the following form for the
operators ${\ha {\car}_{ns_n}}$, 
$\ha A_1$ and $\ha Q$: 
\begin{eqnarray}
 {\ha \car }_{n m s_{nm}} &=& (P^\da)^n  \le[
\theta(m) 
(\hat q^\da)^m + \theta(-m) (\hat {\bar q}^\da)^{-m}
\ri]  {\ha g}_{s_{nm}}\\
\ha A_1 &=& (\ha P^ \da \ha P + 1/2)\bar a~, \la{a1p}
\\
\ha Q &=& e({\hat q}^\da \hat q-\hat {\bar q}^{\da} 
\hat {\bar q})
\la{defn1}
\end{eqnarray}
where $\ha P^\da ~(\ha P)$ raises (lowers) the $A_1$ 
eigen-levels from $n$ to $n+1$ ($n+1$ to $n$). $\hat
q^\da$, 
and $\hat q$ are the usual charge raising
and lowering operators for particle states (i.e.
positive
charge states) 
and $\hat {\bar q}^\dag$,  $\hat {\bar q}$ correspond
to charge raising and lowering of antiparticle
states (negative charge states). The hermitian internal 
operator ${\ha g}_{s_{nm}}$, similar to secret operator
in \cite {gour}, 
transforms the internal quantum state within the  
same $\hat A_1$ eigenstate $n$ and charge eigenstate
$m$. Next, we postulate that these operators
satisfy the following commutation relations such that
(\ref{cl2}-\ref{al3}) 
are satisfied :
\ba
[\ha P, \ha P^\da] &=&[\ha q, \ha q^\da]~=~[\ha {\bar
q}, \ha {\bar q}^\da]~=~1
\la {pr1}\\
{[}\ha q, \ha {\bar q}] &=&[\ha q, \ha {\bar
q}^\da]~=
%~[\ha P^\da, \ha q]
~=0~, 
\la {pa1}\\
{[}\ha P, \ha g_{s_{nm}}]&=&[\ha P^\da, \ha
g_{s_{nm}}]~=~ 
[\ha q, \ha g_{s_{nm}}]~=~0~, \la {pr2}
%[\ha q^\da, \ha g_{s_{nm}}]~=0~, \la {pr2}\\
%{[}\ha {\bar q}, \ha g_{s_m}]&=&[\ha {\bar q}^\da,
%\ha g_{s_m}]=0~, \la {ps2}\\
%{[}\ha q, \ha P]&=&[\ha {\bar q}, \ha g_{s_m}]=0~,
\la {ps2} \\
{[}{\hat g}_{s_{n'm'}},{\hat g}_{s_{nm}}] &=&
\epsilon_{nn'mm'}^{n''m''} 
{\hat g}_{s_{n''m''}} 
~~~~{\rm where}~ \epsilon_{nn'mm'}^{n''m''} \neq 0
~{\rm iff}
~ %s_{n''m''} = s_{nm}+s_{n'm'}~. \la {pr3}  
n'' = n + n' {\rm and} ~m''=m+m'~. \la {pr3}
\ea
Also, the area creation (annihilation) operators 
${\hat P}^\dagger$ (${\hat P}^\dagger$) commute with 
the charge creation (annihilation) operators 
${\hat q}, {\hat {\bar q}}$ (${\hat q}^\dagger, 
{\hat {\bar q}}^\dagger$) and ${\hat g}_{s_{nm}}$
commutes with all other operators. 
Equation (\ref {pr3}) ensures the validity of 
Eq.(\ref {al3'}); however it should
be remembered that the operators $\ha g_{s_{nm}}$ have
a meaning
only within the product form $(\hat P^\da)^n
[\theta(m)(\hat q^\da)^m + 
\theta(-m)(\hat {\bar q}^\da)^{-m}]g_{s_{nm}}$.
Comparison with the operators of reduced phase space
approach (\ref {op1}) shows
that the form of $\ha P^\da$ should be as follows:
\be
{\hat P}^\da = \f{1}{\sqrt{2} } \le[ {\hat X} - i
{\hat \P_X}  \ri]~~.
\la{map1}
\ee
Thus we have an explicit form for $\hat P^\da$ in
terms
of canonically conjugate variables $(X, \Pi_X)$ in
reduced
phase space approach. 
Note that since observables in the reduced phase space approach 
consist of macroscopic quantities like $M$ and $Q$ alone, 
the operators 
$\hat q$ and $\hat {\bar q}$ are `hidden' just as 
the operator $g_{s_{nm}}$. This is perfectly consistent with
the well known fact that the same eigenvalue of $\hat Q$ 
can be obtained in many possible ways as the sum of 
particle $\hat q^\da \hat q$ and antiparticle $\hat
{\bar q}^\da
\hat {\bar q}$ charges. Equivalently, 
from the point of view of an asympotic observer,
the microscopic details of the
particle-antiparticle
charge composition for a given charge state are unobservable.  
Similarly, the microscopic quantum state determined by
the secret operator $\ha g_{s_{nm}}$ cannot be
accessible to the asymptotic observer. These 
arguments are equivalent to the no hair theorem.

We see that approaches I and II are
equivalent in the spherically symmetric sector from
the asymptotic observer viewpoint. 
The algebra studied by Bekenstein is similar to the
problems of single
particle quantum mechanics where non-trivial zero
point energy always
exists except for a free particle. Hence it is not
surprising that the 
vacuum area is non-zero. However note that the precise
value of the
remnant (as well as the quantum of area) remains 
undetermined in this approach. In the
reduced phase space
approach on the other hand, the remnant (and area quantum) 
is explicitly determined
to be a multiple of the Planck area in the relevant
dimension. Since the
latter is the only natural length scale in quantum
gravity, this seems
satisfactory. Moreover, this comparison makes the physical
significance of the abstract operators of Approach II clear. 
They are simply the canonically transformed version of the 
macroscopic gravitational degrees of freedom. This significance 
has also been recently emphasised by Gour \cite{gour2}. 
Finally, note that the discrete area
spectrum (\ref{rnspec})
means that Hawking radiation would consist of discrete
spectrum
lines, enveloped by the semi-classical Planckian
distribution.
As argued in \cite{bek,bk,bdk}, for Schwarzschild
black holes of
mass $M$, the gap is order $1/M$, which is comparable
to the frequency
at which the peak of the Planckian distribution takes
place. Hence the
spectrum would be far from continuum, and can
potentially be tested if and
when Hawking radiation becomes experimentally
measurable.
It would also be interesting to explore the
implications of the
Planck size remnant to the problem of information
loss, since
the presence of the former can considerably influence
black hole
evolution near its end stage \cite{remnants}.

A further test of the correspondence elucidated in
this article would
be to apply it to axi-symmetric rotating black holes.
However, for this,
one has to first extend the reduced phase space
formalism to situations
involving angular momentum, since the former has not
been
explored beyond the realm of spherical symmetry. 

\vs{.4cm}
\no
{\bf Acknowledgements}

S.D. would like to thank A. Barvinsky and G. Kunstatter 
for numerous fruitful discussions, correspondence,
and for collaboration, on which papers \cite{bdk} are based.
S.D. would also like to thank 
J. D. Bekenstein, A. Dasgupta, G. Gour and V. Husain for 
several important comments and criticisms, which helped 
in improving the manuscript. This work was supported in part by 
the Natural Sciences and Engineering Research Council
of Canada.

\end{document}